\documentclass[12pt,aps,prb,preprint]{revtex4}
\usepackage{graphicx}
\usepackage{color}
\usepackage{verbatim}
\usepackage{amssymb}
\usepackage{amsbsy}
\usepackage{amsmath}
\usepackage{bm}
\usepackage{bm}

\begin{document}

\title{Maxwell's Equations in a Uniformly Rotating Dielectric Medium\\ and the Wilson-Wilson Experiment}
\author{C E S Canovan}
  \email{c.canovan1@lancaster.ac.uk}   
\author{Robin W Tucker}
\email{r.tucker@lancaster.ac.uk}
\affiliation{Department of Physics, Lancaster University,    Lancaster, UK and The Cockcroft Institute, Daresbury Laboratory, WA4 4AD, UK}

\date{\today}

\begin{abstract}
This note offers a conceptually straightforward and efficient way to formulate and solve problems in the electromagnetics of moving media based on a representation of Maxwell's equations in terms of differential forms on spacetime together with  junction conditions  at moving interfaces. This framework is used to address a number of issues that have been discussed recently in this journal  about the theoretical description underlying the interpretation of the Wilson-Wilson experiment.
\end{abstract}
\maketitle

\def\iv{i_V}
\def\iu{i_U}
\def\dualv{{\dual{V}}}
\def\dualu{{\dual{U}}}
\def\Real{{\mathbb R}}
\def\Cmpx{{\mathbb C}}
\def\man{{M}}
\def\Set#1{{\left\{#1\right\}}}
\def\dual#1{{\widetilde{#1}}}
\def\GamTM{{\Gamma T\man}}
\def\GamLamM#1{{{\cal S} \Lambda^{#1}\man}}
\def\qquadand{\qquad\text{and}\qquad}
\def\quadand{\quad\text{and}\quad}
\newcommand\BE[1]{{\begin{equation}#1\end{equation}}}
\newcommand\BAE[1]{{\begin{equation}{\begin{aligned}#1\end{aligned}}\end{equation}}}
\newcommand{\GX}{{\cal G}_X}
\newcommand{\EH}{\hat\epsilon}
\newcommand{\MH}{\hat\mu^{-1}}
\newcommand{\lam}{\lambda}
\newcommand{\E}{\epsilon}
\newcommand{\M}{\mu^{-1}}
\newcommand{\T}[1]{\tau^{#1}}
\newcommand{\pr}{\prime}
\newcommand{\CHI}{{\hat \chi}}
\newcommand{\Et}{\epsilon_t}
\newcommand{\Mt}{\mu^{-1}_t}
\newcommand{\gt}{{g_t}}
\newcommand{\git}{{{g_t}^{-1}}}
\newcommand{\bfL}{{\bf \Lambda}}
\newcommand{\Vt}{V_{{\gt}}}
\newcommand{\Pt}{\Phi_{{\gt}}}
\newcommand{\be}[1]{{\begin{equation}#1\end{equation}}}
\newcommand{\tl}[1]{\tilde #1}
\newcommand{\PPi}{{\cal P}}
\newcommand{\wdd}{\wedge}
\def\dotLAM#1{\dot{\Lambda}^{{#1}}}
\def\ivF{\iv F}
\def\ivstarF{\iv{{\star}}F}
\def\iuF{\iu F}
\def\iustarF{\iu{{\star}}F}
\def\ivG{\iv G}
\def\ivstarG{\iv{{\star}}G}
\def\iuG{\iu G}
\def\iustarG{\iu{{\star}}G}
\newcommand{\bfe}{ \TYPE 1    {\mathbf e}  {}   }
\newcommand{\bfb}{ \TYPE 1 {\mathbf b}   {}}
\newcommand{\bfd}{ \TYPE 1 {\mathbf d}   {} }
\newcommand{\bfh}{\TYPE 1  {\mathbf h}   {} }
\newcommand{\acc}{{\cal A}}
\newcommand{\bfE}{ \TYPE 2    {\mathbf E}  {}   }
\newcommand{\bfB}{ \TYPE 2 {\mathbf B}   {}}
\newcommand{\bfD}{ \TYPE 2 {\mathbf D}   {} }
\newcommand{\bfH}{\TYPE 2  {\mathbf H}   {} }
\newcommand\bbfJ{{\bf J}}
\newcommand\bbfj{{\bf j}}
\newcommand{\bfEdot}{ \TYPE 2    {\dot{\mathbf E}}  {}   }
\newcommand{\bfBdot}{ \TYPE 2 {\dot{\mathbf B}}   {}}
\newcommand{\bfDdot}{ \TYPE 2 {\dot{\mathbf D}}   {} }
\newcommand{\bfHdot}{\TYPE 2  {\dot{\mathbf H}}   {} }

\def\Me{{\mathbf e}}
\def\Mb{{\mathbf b}}
\def\Md{{\mathbf d}}
\def\Mh{{\mathbf h}}

\newcommand\TYPE[3]{ \underset {(#1)}{\overset{{#3}}{#2}}  }
\newcommand\eeck[1]{\TYPE #1 {\bfe}  {\circ}}
\newcommand\EP{\epsilon_r\epsilon_0\,}
\newcommand\MU{\mu_r\mu_0\,}
\newcommand\ep{\epsilon}
\newcommand{\hash}{\#}
\newcommand{\hashat}{\hat{\#}}
\newcommand{\xibar}{\underline\xi}
\newcommand{\D}{{\boldsymbol d}\,}
\newcommand{\DHAT}{\hat{\D}\,}
\newcommand{\ot}{\otimes}
\newcommand\PD[1]{\frac{\partial}{\partial #1}}
\newcommand\eLONG{\underset{(0)}  {\hat{\mathbf e}}}
\newcommand\ePERP{\underset{(1)}  {\hat{\mathbf e}}}
\newcommand\hLONG{\underset{(0)}  {\hat{\mathbf h}}}
\newcommand\hPERP{\underset{(1)}  {\hat{\mathbf h}}}
\newcommand\eLONGDOT{\underset{(0)}  {\dot{\hat{\mathbf e}}}}
\newcommand\hLONGDOT{\underset{(0)}  {\dot{\hat{\mathbf h}}}}
\newcommand\ePERPDOT{\underset{(1)}  {\dot{\hat{\mathbf e}}}}
\newcommand\hPERPDOT{\underset{(1)}  {\dot{\hat{\mathbf h}}}}
\newcommand\PSI[2]{\Psi_{#1}{#2}}
\newcommand\PHI[2]{\Phi_{#1}{#2}}
\newcommand\eLONGG{\underset{(0)}  {\check{\mathbf e}}}
\newcommand\ePERPP{\underset{(1)}  {\check{\mathbf e}}}
\newcommand\hLONGG{\underset{(0)}  {\check{\mathbf h}}}
\newcommand\hPERPP{\underset{(1)}  {\check{\mathbf h}}}
\newcommand\VV[2]{V_{#1}^{#2}}
\newcommand\VVV[3]{V_{#1}^{#2(#3)}}
\newcommand\VVVDOT[3]{{\dot V}_{#1}^{#2(#3)}}
\newcommand\II[2]{I_{#1}^{#2}}
\newcommand\III[3]{I_{#1}^{#2(#3)}}
\newcommand\IIIDOT[3]{{\dot I}_{#1}^{#2(#3)}}
\newcommand{\DD}{{\cal U}}
\newcommand\CC[2]{\gamma_{#1}^{#2}}
\newcommand\ccc[3]{\gamma_{#1}^{#2(#3)}}
\newcommand\cccdot[3]{{\dot \gamma}_{#1}^{#2(#3)}}
\newcommand\GGGDOT[3]{{\dot \gamma}_{#1}^{#2(#3)}}
\newcommand\NN[1]{{\cal N}_{#1}}
\newcommand\MM[1]{{\cal M}_{#1}}
\newcommand{\jLONG}{\TYPE 0 {\hat J} {}}
\newcommand{\jPERP}{\TYPE 1 {\hat J} {}}
\newcommand\RHO{\underset{(0)}\rho}
\newcommand\zzz{\left(\frac{\omega}{v} r\right)}
\newcommand\zz{\frac{\omega}{v} r}
\newcommand\ZZ{{\cal Z}_l}
\newcommand\ZZZ{{\cal Z}_L(\zz)}
\newcommand\xmp{x_{mp}}
\newcommand\hp{\hat\chi}
\newcommand\hatchi[1]{\hat\chi(#1)}
\newcommand\zzzz{\zeta}
\newcommand\CCC[2]{{\cal C}_{#1}{}^{#2}{(\zzzz)}}
\newcommand\JJ[1]{J_{#1}\left(\xmp \frac{r}{a} \right)}
\newcommand\union\cup
\newcommand\starr{\breve{\hash}}
\newcommand\z{\zeta}
\newcommand\e{f}
\newcommand\eb{\breve{e}}
\newcommand\stcds{(t,z,\rho,\phi)}
\newcommand\newcds{(t,\z,r,\theta)}
\newcommand\epcds{(\ep,t,\z)}
\newcommand\EPS{{\mu {\cal Y}^2}}
\newcommand\PP{{}^\prime}
\newcommand\vol{\,r\, \D r \wedge \D\theta}
\newcommand\intt{\int_{\DD}}
\newcommand{\EM}{ electromagnetic }
\newcommand{\beq}{\begin{equation}}
\newcommand{\beqa}{\begin{eqnarray}}
\newcommand{\eeq}{\end{equation}}
\newcommand{\eeqa}{\end{eqnarray}}
\newcommand{\non}{\nonumber}
\newcommand{\fr}[1]{(\ref{#1})}
\newcommand{\cj}{\mbox{c.c.}}
\newcommand{\ts}{\widetilde s}
%
\def\ee{\,\epsilon_0\,}
\def\cc{\, c \,}



\vskip0.5cm

\begin{flushleft}{\bf I. INTRODUCTION}\vskip0.5cm\end{flushleft}

The Wilson-Wilson \cite{Wilson} experiment of 1913 is regarded as a classic confirmation  of special relativistic electrodynamics and is thought to have been pivotal in the acceptance and development of the Minkowskian theory of electrodynamics of moving media. In the experiment, an uncharged dielectric cylinder underwent  uniform rotation in a uniform static magnetic field directed along its axis of symmetry. A radially induced electric field was observed whose magnitude depended on the rotation speed,  dielectric constant and permeability of the cylinder.

However in  1990 Pellegrini and Swift \cite{Pell}  published a controversial paper which expressed doubts about the application of special relativity  to the theory behind the experiment. Since then a number of contributors have addressed this issue with arguments of varying degrees of complexity and success.

This paper offers a conceptually straightforward and unambiguous approach to the general problem of finding the electromagnetic fields in a  \lq\lq simple homogeneous and isotropic"  uniformly rotating dielectric  in a static uniform external field by using the  covariant Maxwell equations on spacetime and the covariant induced jump conditions that have to be satisfied at its interfaces with other media, including the vacuum.  It is  concluded that while the results of the original theoretical analysis predicting the observed radial electric field in the Wilson-Wilson experiment are indeed correct, the approach via the use of Lorentz transformations between uniformly moving reference frames can be perplexing. However by formulating electrodynamics in terms of spacetime tensors and exploiting the powerful techniques  based on the calculus of differential forms one circumvents many of the worries that seem to have clouded the debate initiated in Ref. 2. One then seeks solutions satisfying (moving) boundary conditions for media subject to  Minkowski constitutive relations. If these natural relations are accepted there is little else to argue about.

Although the techniques to be outlined can be applied generally the strategy for finding {\it exact} solutions for the fields inside uniformly rotating media in static uniform fields  often relies on exploiting simple geometries for the dielectric domains or expanding solutions as a perturbation in powers of $v/c$ where $v$ is a maximum speed of the periphery of the domain. However for an infinitely long rigid homogeneous dielectric cylinder in an external magnetic field one may find an ${\it exact}$ (relativistic) solution. Although the Wilson-Wilson cylinder had a finite length the effects of the cylinder's ends on the observed potential difference were presumably negligible
 since  in the non-relativistic limit the exact solution reduces to this observed value.

By comparison in the R\"{o}ntgen and Eichenwald \cite{RE} experiment  a much shorter dielectric cylinder (disc) was rotated in a uniform static {\it electric} field directed along the axis of rotation and an induced magnetic field was detected. In order to demonstrate the solution strategy in this situation we consider instead a uniformly rotating dielectric sphere since this has a single smooth interface instead of the three interfaces of the rotating disc of finite thickness.

Section II briefly outlines the history leading to the debate about the theoretical description of the Wilson-Wilson experiment. Section III presents the covariant Maxwell system and associated interface conditions in the language of differential forms and the vector notation of Gibbs for comparison.  In Section IV  these equations are used to analyze  electromagnetic fields inside and outside an infinitely long  uniformly rotating cylindrical shell in an external uniform static magnetic field and the results compared with the   Wilson-Wilson experiment.  In Section V  the same strategy is applied to the problem  of finding the  electromagnetic fields inside and outside a uniformly rotating sphere in an external uniform static electric field. Section VI presents our conclusions.

\vskip0.5cm\begin{flushleft}{\bf II. THE WILSON-WILSON EXPERIMENT AND THE DEBATE SURROUNDING ITS VALIDITY}\vskip0.5cm\end{flushleft}
In 1908 Einstein and Laub \cite{Einstein} used the new theory of special relativity to predict the  fields induced in a dielectric slab moving in a straight line in an external, constant electric field. In a footnote, the authors noted that the theory could be extended to a  material with a cylindrical shape. In 1913 Wilson and Wilson \cite{Wilson} performed their classic experiment in which a dielectric cylindrical shell, of internal and external radii $r_1$ and $r_2$ respectively, rotated with constant  angular speed $\Omega$ about its axis of symmetry in a uniform static magnetic field directed along its axis. A diagram of the set-up is shown in Figure 1.


They then measured the potential difference $V_{12}$  across the inner and outer radii. Following from the calculation in Ref. 4 for the dielectric slab, they argued that (in modern SI units):
\begin{equation}
\label{eq:1}
V_{12} =\mu_r\,(1-\frac{1}{\mu_r\epsilon_r})\,\frac{\Omega}{2}B_0 (r^2_2-r^2_1)
\end{equation}
where $\epsilon_r$ and $\mu_r$ are the relative permittivity and permeability of the cylinder and $B_0$ is the magnitude of the external magnetic field. The magnitude of the radial electric field  generating Eq.~\eqref{eq:1}  is
\begin{equation}
{\vert\bf e}\vert = \mu_r\, ( 1-\frac{1}{\mu_r \epsilon_r})\, r \Omega B_0
\end{equation}
in conformity with the observed radial potential difference.
However doubts were raised  in Ref. 2 about the methodology of  extending the results for the linearly moving dielectric slab to a situation where the medium was not in translational motion. The argument hinged about extending the use of the Lorentz transformation formulae that related fields in different inertial frames  to rotating systems that were only considered to be \lq\lq instantaneously at rest" in an inertial frame.
 Instead  an alternative approach using rotating frames of reference was proposed  that predicted the magnitude of the induced radial electric field to be:
\begin{equation}
\vert{\bf e}\vert = \mu_r ( \frac{1}{\epsilon_r} - 1) r \Omega B_0
\end{equation}
which clearly differs from the result obtained by Wilson and Wilson.
The authors  also applied their arguments to the established predictions of the  R\"{o}ntgen and Eichenwald experiment but were unable to reconcile their results with both these and the Wilson-Wilson prediction.
 The paper provoked a lengthy debate about how constitutive relations in rotating frames of reference should be applied in these contexts.  Hertzberg et al \cite{Hertzberg}  repeated the Wilson-Wilson  experiment and found agreement with the original results, thereby concluding the reasoning in Ref.2 was false.

To place this note in context it is useful to briefly review some of the arguments that  have  contributed to the debate.
Burrows \cite{Burrows} believed that the key issue was whether the constitutive equations  took account of the  \lq\lq non-orthogonal nature" of coordinate systems (his nomenclature).
Weber \cite{Weber}  produced a result agreeing with Wilson and  Wilson,   using  sources and co-ordinate transformations.
Ridgely \cite{Ridgely} used the electromagnetic jump conditions to calculate the electric field in the rotating cylinder at any interface by assuming that it was radially directed. It is worth noting, however, that the direction and dependence of the field on radius was not determined from the Maxwell equations.

An   account of different approaches to the analysis of the Wilson-Wilson prediction has also been produced by McDonald. \cite{McDonald}  Here the author performs a naive calculation of the  electric field induced in a slab of dielectric in uniform rectilinear motion in  a static magnetic field based on pre-special relativistic notions. He notes that this yields erroneous results when compared with results deduced from the use of special relativistic Lorentz transformations and simple constitutive relations. However he also suggests that an analysis based on general relativity should be used to understand how to transfer such an analysis to the Wilson-Wilson rotating shell case. It appears that by this he means the use of tensor methods on spacetime since the need to include Einstein gravitation is unwarranted.  His use of \lq\lq local inertial frames\rq\rq is then used to justify the concordance between the results obtained for the linearly and rotating dielectric by an appeal to the \lq\lq electrodynamics in a rotating frame using general relativity\rq\rq. \cite{Footnote1} We find this approach somewhat contrived in as much that in order to achieve this concordance it is necessary to employ field conditions that ultimately follow from solving Maxwell's equations in the {\it inertial} (laboratory) frame subject to boundary conditions at the interface with the rotating dielectric.

In Hehl\cite{Hehl} (p.359) the analysis of the Wilson-Wilson and R\"{o}ntgen-Eichenwald  experiments is  approached from solutions associated with media in uniform translational motion. The authors then provide a solution in the language of forms that coincides with that derived from our analysis below. Here we explicitly show how  the boundary conditions for media in
rotary motion play a role in deriving such a solution from Maxwell's equations.

In this paper it is argued that many of the conceptual issues about the use of non-inertial frames in special relativity are really irrelevant if one approaches the electromagnetics of a moving medium in terms of tensor fields on spacetime. The potential difficulties reside, not in transforming electromagnetic fields  between different frames of reference, but in finding techniques to solve what are often difficult boundary value problems in the laboratory frame, and being fully convinced about the validity of the assumed constitutive properties of the medium when it is at rest in any inertial frame.

The approach below circumvents these issues by working with a pair of $2-$forms, $F$ and $G$ on spacetime, the relation between them in the vacuum and inside an arbitrarily moving homogeneous, isotropic, non-dispersive medium and their orthogonal decompositions relative to local  observers. Besides being computationally efficient this   allows one to address in what sense it is acceptable to consider whether  points in a rotating medium can be regarded as instantaneously at rest in some local inertial frame of reference.


\vskip0.5cm\begin{flushleft}{\bf III. ELECTROMAGNETIC FIELDS IN SPACETIME}\vskip0.5cm\end{flushleft}

\def\LAM{{\Lambda}}
\def\bfNU{{\mathbf N}^U}
\def\bfNU{\widetilde{\boldmath N}^U}
\def\Ut{\tilde U}
\def\bfvU{{\mathbf v}^U}

\def\dphi{d\,\Phi}
\def\KK{{\cal K}}
\def\bflU{{\mathbf l}^U}
\def\bfSU{{\mathbf S}^U}
\def\bfsU{{\mathbf s}^U}
\def\bfkU{{\boldsymbol \kappa}^U}
\def\bfsigmaU{{\boldsymbol \sigma}^U}
\def\wt{\widetilde}
\def\dpp{{d\Phi \cdot d\Phi}}
\def\bfa{{ \boldsymbol \alpha }^U\,}
\def\bfbU{{ \boldsymbol \beta }^U\,}
\def\UN{{}}
\def\vn{{ v}_{N^U}\,}
\def\sig0{{\sigma_0^U}}
\def\bfvNU{{\vn}}

 The language of differential forms  has been  employed a number of times in this journal; see for example  the  papers by Amar,\cite{Amar} Hoshino \cite{Hoshino} and  Schleifer. \cite{Schleifer} Introductory  books on the subject include Flanders, \cite{Flanders} Burke, \cite{Burke} Darling \cite{Darling} and Benn \& Tucker, \cite{BT} all of which have sections on differential geometry and electromagnetism.

Maxwell's equations for an electromagnetic field in an arbitrary
medium can be written in terms of  exterior derivatives and differential forms on spacetime:
\begin{align}
d\,F=0 \qquadand d\,\star\, G =j \label{eq:4}
\end{align}
where $F$ is the Maxwell $2$-form, $G$ is
the excitation $2$-form and $j$ is the $3$-form electric
$4$-current source\cite{Footnote2} on a spacetime  endowed with a metric tensor field $g$ (signature $(-1,+1,+1,+1) $) and associated Hodge dual map $\star$.
The electric $4$-current $j$  describes both  (mobile) electric charge
and effective (Ohmic) currents in a conducting medium.
 To close this system, ``electromagnetic constitutive
relations'' relating $G$ and $j$ to $F$ are necessary. It is often convenient to write $F=d\,A$ (locally) in terms of a $1-$form potential $A$ on spacetime since this automatically solves the first $3-$form equation in Eq.~(\ref{eq:4}).

A  {\it unit} future-pointing timelike
$4$-velocity vector field $U$ on spacetime  may be used to describe an {\it observer frame}.
 The {\it
electric }  $1-$form field $\Me^U$ and {\it magnetic induction    $1-$form
field $\Mb^U$} associated with $F$   in the frame  $U$ are defined by
\begin{align}
\Me^U = \iuF \qquadand \cc\Mb^U = \iustarF \label{intro_e_b}.
\end{align}
Thus, in terms of the contraction operator $i_U$,  $i_U\Me^U=0$ and $i_U\Mb^U=0$ and since $g(U,U)=-1$, one has the orthogonal decomposition of $F$ relative to $U$:
\begin{equation}
F=\Me^U\wedge \dualu - \star\,(\cc\Mb^U\wedge \dualu) \label{intro_F}.
\end{equation}
The $1-$form $\dualu$ is related directly to the vector field $U$ by the metric and defined by the relation $\dualu(X)=g(U,X)$ for all vector fields  $X$.

The {\it displacement}   $1-$form field $\Md^U$ and the
{\it magnetic}   $1-$form field $\Mh^U$  associated with $G$ are
defined with respect to $U$ by
\begin{align}
\Md^U = \iu G\,, \qquadand \Mh^U/\cc = \iu\star G\, \label{Media_d_h}.
\end{align}
Thus
\begin{equation}
G=\Md^U\wedge \dualu - \star\,((\Mh^U/\cc)\wedge \dualu) \label{Media_G}
\end{equation}
and $i_U\Md^U=0$ and $i_U\Mh^U=0$.
It follows from these definitions that
\begin{align}
\star F = c\Mh^U \wedge \dualu + {\boldsymbol E}^U\label{MediaSTARF}
\end{align}
\begin{align}
\star G = \frac{\Mh^U}{c} \wedge \dualu + {\boldsymbol D}^U\label{MediaSTARG}
\end{align}
where ${\boldsymbol E}^U=\star(\Me \wedge \dualu)$  and ${\boldsymbol D}^U= \star(\Md^U \wedge \dualu) $.
The arbitrary motion of a spatially compact medium can be described by a {\it unit} timelike (future-pointing) $4$-velocity vector field $V$ on the history of the medium in spacetime.
A simple non-conducting, non-dispersive isotropic medium can be characterized by the scalar field permittivity $\epsilon=\epsilon_r\,\epsilon_0$ and scalar field permeability $\mu=\mu_r\,\mu_0$ and the constitutive relations
$ \Mh^V= \Mb^V/\mu$ and $\Md^V=\epsilon \,\Me^V  $. In terms of $F$, $G$ and $V$ these relations take the covariant form:
\begin{equation}
\label{eq:11}
\frac{G}{\epsilon_0}=\left(\epsilon_r-\frac{1}{\mu_r}\right) i_V F \wedge \widetilde{V}+\frac{1}{\mu_r}F.
\end{equation}

A vector field  in spacetime assigns a direction (tangent vector) to each event in spacetime. Curves in spacetime  with any of these tangent vectors constitute the integral curves of the vector field.
Integral curves of the frame field $U$, with tangents pointing to the future, model the histories of (idealized) observers. The field $U$ describes a global inertial frame if its components are constant in a global coordinate system (Cartesian coordinates) in Minkowski spacetime. Integral curves of the time-like $4-$velocity field $V$, associated with some material continuum, model the histories of point material constituents.\cite{Footnote3} If $U\neq  V$ but at some event $p$ in spacetime their integral curves share the same
tangent then it is sometimes said that $V$ is {\it instantaneously at rest at $p$ with respect
to the timelike frame $U$}.  Clearly not all points in a rotating medium can be instantaneously at rest with the same event in any inertial frame. These mathematical idealizations enable one to understand the physical assumptions behind the particular constitutive relation in Eq.~(\ref{eq:11}).

In the introductory section of Ref. 5,
 in connection with the Wilson-Wilson experiment one reads  \lq\lq Their predictions implicitly assume that each point in the
cylinder can be treated as if it were in a locally inertial reference frame instantaneously co-moving
 with the point\rq\rq. One may paraphrase this by saying that it is equivalent to assuming that electromagnetic interactions at the molecular and atomic level depend on the velocity of charges and not their accelerations. It also assumes that clocks used by relatively moving observers measure individual {\it proper time} and are not influenced by the possible acceleration of the clock.  The behaviour of real atomic clocks conforms well to this assumption and the standard minimal coupling of charge to the vector potential in quantum electrodynamics is consistent with the former assumption.  Clearly one may contemplate the significance of acceleration dependent interactions involving radiation reaction or intense  gravitational fields  that may bring into question such assumptions. However such modifications appear irrelevant in the context of the experiments under discussion.

The polarisation $2-$form $\Pi$ is defined by $\Pi=G-\epsilon_0 F$. One can then write the second equation in Eq.~(\ref{eq:4}) as
\BE{ \epsilon_0 d \star F= j + \widehat j} where the {\it bound $4-$current $3-$form}
\BE{\widehat j\equiv  -d\star \Pi = {\widehat {\bf J}}^U \wedge\dualu + {\widehat{\boldsymbol\rho}}^U\label{eq:13}}
in terms of {\it bound current $2-$forms  $ {\widehat {\bf J}}^U  $   and  bound charge $3-$ forms $  {\widehat{\boldsymbol\rho}}^U  $   } and \BE{ \Pi = {\bf p}^U\wedge \dual U - \star\left( \frac{{\bf m}^U}{c}\wedge \dual U \right) } in terms of the polarisation $1-$forms ${\bf p}^U$ and  magnetisation $1-$forms ${\bf m}^U$ in the frame $U$.

A moving interface between a medium and the vacuum in space  gives rise to a hypersurface $\Phi=0$ in spacetime describing the interface history of the medium. If there are no free sources, $j=0$, then the Maxwell equations above imply:
\BE{ [F] \wedge d\Phi\vert_{\Phi=0}=0\label{eq:15}}
\BE{ [\star \,G] \wedge d\Phi\vert_{\Phi=0}=0\label{eq:16}}
in terms of the discontinuities $[F]$ and $[G]$ across the history of the interface.
These covariant junction conditions  induce relations among the jumps  $[\Me^U], [\Md^U], [\Mb^U],[\Mh^U]$
across any interface in space with instantaneous Newtonian $3-$velocity ${\bf v}$ in the frame $U$.
In Gibbs' $3-$vector notation these relations become  $$ {\bold N}^U \cdot [{\bold d}^U]= 0 $$
\BE{ \vn [ {\bold d}^U ] + ( {\bold N}^U \times [ {\bold h}^U ] ) = 0 }
\BE{ {\bold N}^U \cdot [ \bf b^U ]=0 }
\BE{ \vn [ {\bold b}^U ] - ( {\bold N}^U \times [ {\bold e}^U ] ) = 0 }
at the interface with    unit normal $ {\bf N}^U $ and $\vn= {\bf v} \cdot {\bf N}^U  $   the normal component of the $3-$velocity ${\bf v}$ there.

In general  finding  solutions of Eq.~(\ref{eq:4}) in different domains with different constitutive properties, that match across moving interfaces in some frame according to the junction conditions Eqs.~(\ref{eq:15}) and (\ref{eq:16}) and satisfy appropriate asymptotic boundary conditions at spatial infinity, is a non-trivial problem, particularly if the spatial boundaries of the media are neither simple nor smooth. The tractable text-book  situation  arises for a compact, electrically neutral, simple polarisable or magnetizable body inserted at rest into a uniform static electric or magnetic field. In that case one expresses the fields outside the medium as  a superposition of the external field and a sum of static (electric or magnetic) multi-pole solutions that can be matched to satisfy the junction conditions across a single interface to a suitable static non-singular Maxwell solution inside the medium. The interior solution can often be determined from a suitable ansatz for the fields or potentials exploiting any symmetry in the problem. Once such a solution is determined it may be shown to be unique.

This strategy can be extended to the situation where such a body is given a uniform rotation about a fixed axis in space, as will be demonstrated below for a infinitely long uncharged dielectric cylindrical shell rotating about its axis of symmetry in an external uniform static magnetic field along this axis, and an uncharged  dielectric sphere uniformly rotating about the direction of a uniform static electric field.

\vskip0.5cm\begin{flushleft}{\bf IV. MAXWELL'S EQUATIONS AND THE WILSON-WILSON PREDICTION}\vskip0.5cm\end{flushleft}
A cylindrical co-ordinate system in {\it flat spacetime} with coordinates  $(t,r,\theta,z)$
is naturally adapted to the problem of a rigidly uniformly (right circular) rotating cylindrical shell of internal and external radii $r_1$ and $r_2$. In these coordinates the metric tensor on spacetime takes the form:
 \begin{equation}
g=-c^2 dt \otimes dt + dr \otimes dr + r^2 d\theta \otimes d\theta +  d z \otimes dz
\end{equation}
and the laboratory reference frame is defined by the unit timelike vector
 \begin{equation}
 U = \frac{1}{c}\frac{\partial}{\partial t}.
\end{equation}
In order to solve this problem consider first an ansatz for  the Maxwell 2-form   $F_{in} $ inside the dielectric, $ r_1 < r < r_2  $:
\begin{equation}
F_{in}=\alpha(r)  dt\wedge dr + \beta(r) dr \wedge d\theta
\end{equation}
where $\alpha(r)$ and $\beta(r)$ are  functions to be determined. The above ansatz  is chosen to yield a stationary radial electric and stationary axial magnetic field in the rotating shell.
 Since $d\,^2=0$, $F_{in}$  immediately satisfies the covariant first Maxwell equation in the system Eq.~(\ref{eq:4}).
For constant  $\Omega< \frac{c}{r_2}$, the cylinder's bulk 4-velocity field is, in these coordinates,
\begin{equation}
V=\frac{\partial_t + \Omega \,\partial_\theta}{\sqrt{c^2- r^2 \Omega^2}}.
\end{equation}

 The excitation two-form in the cylinder follows immediately from Eq.~(\ref{eq:11}):
$$
G_{in}  = \frac{\epsilon_0\left(\alpha(r)(r^2\Omega^2-c^2\epsilon_r\mu_r)+\beta(r)c^2\Omega(\epsilon_r\mu_r-1)\right)}{\mu_r(r^2\Omega^2-c^2)} \hspace{1mm} dt \wedge dr$$
\BE{\qquad   + \frac{\epsilon_0\left(\alpha(r)r^2\Omega(1-\epsilon_r\mu_r)+\beta(r)(r^2\Omega^2\epsilon_r\mu_r-c^2)\right)}{\mu_r(r^2\Omega^2-c^2)} \hspace{1mm}dr \wedge d\theta.
}

This is now substituted into the second equation in Eq.~(\ref{eq:4}) with $j=0$ (since the medium is uncharged) to determine $\alpha(r), \beta(r)$. With $\epsilon_r$ and $\mu_r$ constant, the resulting first order ordinary differential equations  are trivially soluble for $\alpha(r)$ and $\beta(r)$ in terms  of two constants of integration, $C_1$ and $C_2$, giving:
\begin{eqnarray*}
F_{in} = -\frac{(C_1\,+ r^2\,C_2)dt \wedge dr}{r(\Omega^2 r^2 -c^2)}
\end{eqnarray*}
\begin{equation}
+ \frac{(\Omega^2\,C_1\,(r^2\,\epsilon_r\,\mu_r\,\Omega^2 -2\,\epsilon_r\,\mu_r\,c^2+c^2)+c^2\,C_2(r^2\Omega^2-c^2\,\epsilon_r\,\mu_r))r\,dr\wedge\,d\theta}{( \epsilon_r\mu_r -1  ) \Omega \, c^4 (\Omega^2\,r^2 -c^2  )}
\end{equation}
\begin{equation}
G_{in}=\frac{C_1 \,\epsilon_0\,\epsilon_r\,dt\wedge dr}{r\,c^2}+\frac{r\,\epsilon_0\,\epsilon_r\,dr\wedge d\theta(C_1\,\epsilon_r\,\mu_r\,\Omega^2\,+\,C_2\,c^2)}{(\epsilon_r\,\mu_r-1)\,c^4\,\Omega}.
\end{equation}
 Any ansatz for $F_{out}$ in the vacuum regions, $ 0< r < r_1$ and $ r > r_2  $  must include the applied magnetic induction field of constant magnitude $B_0$ and satisfy $d\,F_{out}=0$. Additionally, as there are no free charge sources, we know that there can be no radially-directed electric fields in the vacuum. The exterior ansatz is thus
\begin{equation}
F_{out}= c B_0\, r dr \wedge d\theta.
\end{equation}
In the vacuum $G_{out}=\epsilon_0 F_{out}$ so $d\,\star G_{out}=0$ also.
The outer  and inner interfaces between the rotating cylinder and the vacuum are  the hypersurface $\Phi_2\equiv r-r_2=0$ and
$\Phi_1\equiv r-r_1=0$ respectively.
 The interface conditions Eqs.~(\ref{eq:15}) and (\ref{eq:16}), together with the absence of sources in the vacuum, immediately determine $C_1=0$ and $$C_2=\frac{c^3\,B_0\Omega\,(\epsilon_r\mu_r\,-\,1  )}  {\epsilon_r}. $$
  Thus:
\begin{equation}
F_{in}=\frac{c^3 B_0 r \Omega(1-\epsilon_r\mu_r)dt \wedge dr}{\epsilon_r(r^2\Omega^2-c^2)} + \frac{c B_0 r(r^2\Omega^2-c^2\epsilon_r\mu_r)dr \wedge d\theta}{\epsilon_r(r^2\Omega^2-c^2)}
\end{equation}
\begin{equation}
F_{out}=c B_0 r dr \wedge d\theta
\end{equation}
\begin{equation}
G_{in}=\epsilon_0 c B_0 r dr \wedge d\theta
\end{equation}
\begin{equation}
G_{out}=\epsilon_0 c B_0 r dr \wedge d\theta.
\end{equation}
From these solutions, the $1-$form fields  ${\bf e^U,b^U,d^U} $ and ${\bf h^U}$  inside and outside the cylinder follow as:
\begin{equation}
{\bf e^U}_{in}=\frac{-c^2 B_0 \Omega (\epsilon_r\mu_r -1)r dr}{\epsilon_r\,(r^2\Omega^2 -c^2)}
\end{equation}
\begin{equation}
{\bf b^U}_{in}=\frac{(r^2\Omega^2-\epsilon_r\mu_r c^2)B_0 dz}{\epsilon_r(r^2\Omega^2-c^2)}
\end{equation}
\begin{equation}
{\bf d^U}_{in}=0
\end{equation}
\begin{equation}
{\bf h^U}_{in}=\frac{B_0}{\mu_0}dz
\end{equation}
\begin{equation}
{\bf e^U}_{out}=0
\end{equation}
\begin{equation}
{\bf b^U}_{out}=B_0dz
\end{equation}
\begin{equation}
{\bf d^U}_{out}=0
\end{equation}
\begin{equation}
{\bf h^U}_{out}=\frac{B_0}{\mu_0}dz.
\end{equation}

Since ${\bf p^U}=i_U\Pi$ and $ {\bf m^U}=c\, i_U\star\Pi  $ one  can readily calculate from $\Pi_{in}= G_{in} - \epsilon_0\, F_{in}$ the interior polarisation and magnetisation $1-$forms ${\bf p^U}_{in}$ and ${\bf m^U}_{in}$:

\begin{equation}
{\bf p^U}_{in}=\left(\mu_r-\frac{1}{\epsilon_r}\right)\frac{c^2 \Omega \epsilon_0 B_0 r}{r^2 \Omega^2 -c^2}dr
\end{equation}
\begin{equation}
{\bf m^U}_{in}=\frac{\epsilon_0 c^2 B_0(c^2 \epsilon_r(\mu_r-1) + r^2 \Omega^2(\epsilon_r-1))}{\epsilon_r(r^2 \Omega^2-c^2)}dr
\end{equation}
and from Eq.~(\ref{eq:13}) these give rise to the
bound charge $3-$form
\begin{equation}
\bm{\hat{\rho}^U}=-\frac{2 c^4 \Omega \epsilon_0 B_0 (\epsilon_r\mu_r-1)r}{\epsilon_r(r^2 \Omega^2-c^2)^2}dz \wedge dr \wedge d\theta
\end{equation}
and bound current $2-$form
\begin{equation}
\bm{\widehat{J}^U}=\frac{2 c^3 \epsilon_0 B_0 r \Omega^2(\epsilon_r \mu_r -1)}{\epsilon(r^2 \Omega^2 - c^2)^2}dz \wedge dr
\end{equation}
respectively. The polarisation current $2-$form is equivalent to  an azimuthal directed  current vector field with magnitude $  \bm{\widehat{J}^U}/(dz \wedge dr)$
   while the  scalar charge density induced by the rotation induced polarisation is $\bm{\hat{\rho}^U}/(dz \wedge dr \wedge r\,d\theta)$. The physical origin of the rotationally induced electric and magnetic fields can be associated with these rotationally induced sources.

 In the experiment the velocity of any point with radius $r$ in the cylinder was non-relativistic.  For $r_1 < r < r_2$   the above exact solutions  yield  to leading order in $r\, \Omega/c$
\begin{equation}
{\bf e^U}_{in}\simeq\frac{B_0(\epsilon_r\mu_r-1)r \Omega dr}{\epsilon_r}
\end{equation}
\begin{equation}
{\bf b^U}_{in}\simeq\mu_r B_0 dz.
\end{equation}
The non-relativistic electric field above  is in full agreement with the prediction of Wilson and Wilson.

\vskip0.5cm\begin{flushleft}{\bf V. FURTHER APPLICATIONS}\vskip0.5cm\end{flushleft}

  R\"{o}ntgen and Eichenwald carried out a series of experiments  between 1888 and 1904 involving a thin dielectric disk that was rotated in a uniform static electric field directed along the axis of rotation. To detect  an induced magnetic field various modifications were made to the disc and source of the electric field that make the system  difficult to analyze analytically. Furthermore a simplified system consisting of  a uniformly rotating dielectric disc of finite thickness poses a challenging boundary value problem since in cylindrical coordinates one has interfaces at $z=0$ and $z=z_0$ as well as at $r=r_2$ to consider. The authors are not convinced that a reliable exact analytic solution to this problem exists.

  The case of a  dielectric sphere of radius $a$, uniformly rotating about an axis determined by an external static uniform electric field is more amenable to analysis using Eqs.~(\ref{eq:4}), (\ref{eq:15})  and (\ref{eq:16}). Such a situation is illustrated in Figure 2.


   In this case one naturally works in spherical polar coordinates $\{t,r,\theta,\phi  \}$ in which $g$ takes the form:
 \begin{equation}
g=-c^2 dt \otimes dt + dr \otimes dr + r^2 d\theta \otimes d\theta + r^2 \sin^2\theta d\phi \otimes d\phi.
\end{equation}

Again the general strategy proceeds as before by seeking external solutions to Eq.~(\ref{eq:4}) in terms of a series of static electric and magnetic multipoles in spherical polar coordinates together with the external field and a regular  internal solution that can be matched to the series at $r=a$ using Eqs.~(\ref{eq:15})  and (\ref{eq:16}). Furthermore the complete solution for the fields should reduce to the standard text-book solution for a non-rotating homogeneous isotropic dielectric sphere in a static uniform external field when $\Omega=0$. A perturbative approach in $a\,\Omega/c$ to this type of problem has been suggested by Van Bladel \cite{BLAD} and can be employed here.

Motivated by the solution  for a  non-rotating uncharged dielectric sphere in an external static electric field along the Cartesian $z-$axis, of constant magnitude $E_0$, a natural ansatz for $A_{in}$ and $A_{out}$, to first order in $a\,\Omega/c$, is
\begin{equation}
A_{in}\simeq K_0  r \cos\theta\, dt + \Omega K_1 r^3 \cos \theta\, \sin^2 \theta \,d\phi
\end{equation}
\begin{equation}
A_{out}\simeq E_0 \,r \cos\theta \,dt +\frac{P_0   \cos \theta}{r^2}\,dt +  \frac{ \Omega P_1  \cos \theta \sin^2 \theta }{r^2}\,\,d\phi
\end{equation}
where $K_0$, $K_1$, $P_0$ and $P_1$ are constants independent of $\Omega$.
The first term in $A_{in}$ generates a uniform electric field and the second an interior (non-singular) spherical magneto-static quadrupole field.
The first term in $A_{out}$ generates a uniform electric field, the second  generates an exterior  spherical electrostatic  dipole field and the third term generates  an exterior spherical magnetostatic quadrupole field.
 Inserting these into Eq.~(\ref{eq:4})
   and applying the junction conditions at  $r=a$ yields:  $ K_0=3\,E_0/(\epsilon_r+2) $,
 $K_1=E_0\,(\epsilon_r\mu_r -1)/( c^2 (2+ \epsilon_r  ))$, $P_0 =-E_0\,a^3(\epsilon_r-1)/(2+\epsilon_r) $   and  $ P_1=a^5\, K_1  $. Hence
\begin{equation}
{\bf e}^U_{in} \simeq  \frac{3 E_0\,(\sin\theta\, r \,d\theta - \cos \theta\, dr)}{c\,(\epsilon_r + 2)}
\end{equation}

\begin{equation}
{\bf b_{in}}^U \simeq \frac{E_0\,\Omega \,r \,(\epsilon_r \mu_r -1)}{c^3(\epsilon_r+2)}\left((3 \cos^2 \theta -1)\,dr - 3 \cos \theta \sin \theta \,r\, d \theta\right)
\end{equation}

\begin{equation}
{\bf e}^U_{out} \simeq   \frac{E_0}{c}(r \sin \theta\, d\theta-\cos \theta \,dr)+\frac{E_0 \,a^3 \,(1-\epsilon_r)}{c r^3 (\epsilon_r + 2)}(r \sin \theta \, d\theta+ 2\cos \theta\, dr)
\end{equation}
\begin{equation}
{\bf b}^U_{out} \simeq \frac{E_0 \,\Omega\, a^5 \,(\epsilon_r \mu_r-1)}{c^3(\epsilon_r+2)r^4}\left((3\cos^2 \theta-1)dr+2\cos \theta \sin \theta \,r \,d\theta \right)
\end{equation}
to first order in $a\,\Omega/c$.
The same method can also be used to find an approximate analytic solution for a dielectric sphere uniformly rotating in an external  magnetic field.

\vskip0.5cm\begin{flushleft}{\bf VI. CONCLUSIONS}\vskip0.5cm\end{flushleft}
The  representation of Maxwell's equations in terms of differential forms on spacetime together with  junction conditions  at moving interfaces associated with simple hypersurfaces   offers a conceptually straightforward way to formulate problems in the electromagnetics of moving media.
 This approach has been illustrated by revisiting the debate initiated by the concerns raised by Pellegrini and Swift about the Wilson-Wilson experiment. It is to be noted that the calculation of the basic solutions benefits from working directly with the forms $F, A$ and $G$ on spacetime rather than electric and magnetic fields on space and the bound sources $ \widehat{\bf J}^U, \widehat{\boldsymbol \rho}^U $ encoded in the  polarization form $\Pi$. The structure of the polarization and magnetization sources  can aid one's  physical intuition about the origin of the fields induced by moving media once these solutions  have been obtained.
  Throughout our analysis no direct use has been made of Lorentz transformations between inertial reference frames. Essential use has been made of {\it local reference frames} at each event in spacetime associated with the medium and {\it local frames} associated with observers. In this manner there is no restriction on the physical motion of either local observers or the motion of the medium. For the analysis of the experiments discussed in this note (where gravitational effects are irrelevant)  laboratory observers have been chosen to be globally inertial while the medium has been assigned a (uniform) rotary acceleration.
 The method has also been applied to the computation of the fields induced when a polarisable sphere rotates in a uniform static electric field following a general procedure that can be used to generate solutions  as an  expansion in powers of the rotation speed.

In our view a clarification of the role of special relativity in dealing with the Wilson-Wilson debate offers  several insights for understanding more generally the electromagnetic response of accelerating media.

\newpage

\bibliographystyle{revtex}

\begin{thebibliography}{9}
\bibitem{Footnote2}
All tensors in this article have dimensions
  constructed from the SI dimensions $[M], [L], [T], [Q]$ where $[Q]$
  has the unit of the Coulomb in the MKS system. We adopt $[g]=[L^2],
  [G]=[j]=[Q],\,[F]=[Q]/[\ee]$ where the permittivity of free space
  $\epsilon_0$ has the dimensions $ [ Q^2\,T^2 M^{-1}\,L^{-3}] $ and
  $c=\frac{1}{\sqrt{\ee\mu_0}}$ denotes the speed of light in vacuo.
  Note that, with $[g]=[L^2]$, for $p-$forms $\alpha$ in $n$ dimensions one has $[\star \alpha]=[\alpha] [L^{n-2p}]$
\bibitem{Footnote3}
Where $i_V\,d\tilde V \neq 0$ the medium has an acceleration.
\end{thebibliography}

\end{document}